\documentclass[aps,11pt]{revtex4}
\usepackage[dvips]{graphicx}

\usepackage{amsmath}
\usepackage{bm}

\begin{document}

\preprint{}

\title{How Long Could We Live?}
\author{Ilja Dorsner$^{1}$}
\email{idorsner@ictp.trieste.it}
\author{Pavel Fileviez P\'erez$^{1,2,3}$}
\email{fileviez@cftp.ist.utl.pt}
\affiliation{$^{1}$The Abdus Salam
International Centre for Theoretical Physics\\
Strada Costiera 11, 34014 Trieste, Italy. \\
$^{2}$Pontificia Universidad Cat\'olica de Chile. 
Facultad de F{\'\i}sica, Casilla 306 \\
Santiago 22, Chile.\\
$^{3}$ Centro de F{\'\i}sica Te\'orica de Part{\'\i}culas. 
Departamento de F{\'\i}sica. Instituto Superior T\'ecnico. 
Avenida Rovisco Pais,1\\ 1049-001 Lisboa, Portugal.}
\begin{abstract}
We investigate model independent upper bounds on total proton
lifetime in the context of Grand Unified Theories with the
Standard Model matter content. We find them to be $\tau_p \leq
1.5^{+0.5}_{-0.3} \times 10^{39} \
\frac{(M_X/10^{16}\,\textrm{GeV})^4}{\alpha_{GUT}^2} \
(0.003\,\textrm{GeV}^3 / \alpha)^2\,\textrm{years}$ and $\tau_p
\leq 7.1^{+0.0}_{-0.0} \times 10^{36} \
\frac{(M_X/10^{16}\,\textrm{GeV})^4}{\alpha_{GUT}^2} \
(0.003\,\textrm{GeV}^3 / \alpha)^2\,\textrm{years}$ in the
Majorana and Dirac neutrino case, respectively. These bounds, in
conjunction with experimental limits, put lower limit on the mass
$M_X$ of gauge bosons responsible for the proton and bound-neutron
decay processes. For central values of relevant input parameters
we obtain $M_X \geq 4.3 \times 10^{14}
\sqrt{\alpha_{GUT}}\,\textrm{GeV}$. Our result implies that a
large class of non-supersymmetric Grand Unified models, with
typical values $\alpha_{GUT} \sim 1/39$, still satisfies
experimental constraints on proton lifetime. Our result is
independent on any CP violating phase and the only significant
source of uncertainty is associated with imprecise knowledge of
$\alpha$---the nucleon decay matrix element.
\end{abstract}
\maketitle
\section{Introduction}

Grand Unified Theories~\cite{SU(5),SUSYSU(5),SO(10)1,SO(10)2}
(GUTs) are the most appealing extensions of the Standard Model
(SM) of strong and electroweak interactions. Being founded on the
ideas of force and matter unification they always generate two
predictions regardless of their exact realization; one is the
gauge coupling unification and the other is the proton decay. Of
the two it is the latter that offers the \textit{only}\/
unambiguous way to test GUTs~\cite{PatiSalam}. However, despite
systematic experimental search it has not been observed so
far~\cite{Hirata:1989kn,Shiozawa:1998si,McGrew:1999nd}. Even if it
is observed, a clear test of GUT might prove difficult due to
inherent model dependence of all relevant proton decay
contributions~\cite{Weinberg,Zee,Sakai}. Regardless of that, it is
worth asking whether we can expect the test of the GUT idea
through proton decay experiments with certainty.

There are several generic contributions to nucleon decay in GUTs.
(For an incomplete list of various studies on proton decay
constraints on different unifying theories
see~\cite{Dimopoulos,Buras,Nathp1,Nathp2,Hisano1,Raby,Goto,Murayama,Dermisek}.)
In the non-supersymmetric case the most important ones are the
Higgs and gauge $d=6$ contributions. In supersymmetric theories
there are two more contributions that generically predict too
rapid proton decay. These are the $d=4$ and $d=5$ contributions.
Of course, the four contributions we mention do not encompass all
the possibilities. (For example, presence of extra Higgs
representations such as $\bm{15}$ or $\bm{10}$ in an $SU(5)$ GUT
can lead to additional contributions through mixing of appropriate
components of these representations with the triplet partners of
the usual Higgs doublets~\cite{MohapatraBook,IljaFuture}. It is
also possible to have sizable contributions without any reference
to the GUTs if the theory is supersymmetric~\cite{Murayama:2}.)
But, they are certainly the most generic ones.

It may come as a surprise that despite their multiplicity and
diverse origins all of these contributions can in principle be
completely suppressed or forbidden \textrm{except}\/ the gauge
$d=6$ ones. For example, the so-called matter parity forbids the
dangerous $d=4$ contributions and there are numerous different
ways to efficiently suppress the $d=5$ operators and Higgs $d=6$
operators in realistic scenarios. (For discussion on suppression
of $d=5$ operators see for example~\cite{Bajc1,Bajc2,Costa}.) In
essence, the most promising tests of GUTs can be done through the
gauge $d=6$ contributions.

The idea of using the gauge $d=6$ dominated branching ratios for
the two-body nucleon decays to distinguish between different GUT
models of fermion mass has been introduced in the pioneering work
of De Rujula, Georgi and Glashow~\cite{DeRujula}. Their idea has
been revisited and elaborated on more recently. Namely, it has
been shown that it is possible to make clear test of any GUT with
symmetric Yukawa couplings through the nucleon decay channels into
antineutrinos~\cite{Pavel}. Similar conclusions~\cite{Ilja1} also
hold in the context of flipped
$SU(5)$~\cite{DeRujula,Barr,Derendinger,Antoniadis}. There, the
clear test requires symmetric Yukawa couplings in the down-quark
sector only. (Flipped $SU(5)$ is to be considered a true GUT in
the case of further embedding in $SO(10)$.)

But, in general, even the gauge $d=6$ contributions can be
significantly suppressed if not set to zero. For example, one can
completely rotate them away in the flipped $SU(5)$
context~\cite{Ilja2}. The relevant contributions there, which we
refer to as the ``flipped $SU(5)$ contributions'' for obvious
reason, represent only one half of all possible gauge $d=6$
contributions in GUTs. The other half, which we refer to as the
``$SU(5)$ contributions'' which is due to exchange of proton decay
mediating gauge fields present in an $SU(5)$ gauge group cannot be
rotated away without a conflict with the measurements on fermion
mixing~\cite{Nandi}. Nevertheless, it is worth investigating how
efficiently one can suppress these contributions, too. Since there
are no other gauge $d=6$ contributions besides the two we mention,
this allows us to set an absolute upper bound on nucleon decay
lifetimes. Crucial importance of those bounds lies in the fact
that they are the only way to know if there is ever hope to test
the idea of grand unification with certainty through proton decay
experiments. Even if these bounds turn out to be beyond the
experimental reach they set correct lower limit on $M_{GUT}$
through an absolute lower bound on the mass of the nucleon decay
mediating gauge bosons. In other words, they are the bounds that
can tell us which GUT scenarios are \textit{a priori}\/ ruled out
by experimental data. In what follows we concentrate on GUTs with
the SM matter content, i.e., the three generation case, due to
their phenomenological relevance.

\section{Looking for an upper bound on the total proton lifetime}

To establish an upper bound on the total proton lifetime we first
critically analyze all possible gauge $d=6$ operators contributing
to proton decay. Again, we concentrate solely on these
contributions since all other contributions can be set to zero.

Proton lifetime induced by superheavy gauge boson exchange can be
written as follows
\begin{equation}
\tau_p = C \ M_X^4 \ {\alpha^{-2}_{GUT}} \ m_p^{-5},
\end{equation}
where $C$ is a coefficient which contains all information about
the flavor structure of the theory. $M_X$ is the mass of the
superheavy gauge bosons. $\alpha_{GUT}=g^2_{GUT}/4 \pi$, where
$g_{GUT}$ is the coupling defined at the GUT scale (the scale of
gauge unification). To find a true upper bound on the total
lifetime we need to find the maximal value for the $C$
coefficient. Then, for a given value of $M_X$ and $\alpha_{GUT}$
we can bound the GUT scenario prediction for the nucleon lifetime.

The relevant gauge $d=6$ operators contributing to the decay of
the proton, in the physical basis~\cite{Pavel}, are:
\begin{subequations}
\begin{eqnarray}
\label{Oec} \textit{O}(e_{\alpha}^C, d_{\beta})&=&
c(e^C_{\alpha}, d_{\beta}) \ \epsilon_{ijk} \ \overline{u^C_i} \
\gamma^{\mu} \ u_j \ \overline{e^C_{\alpha}} \
\gamma_{\mu} \ d_{k \beta}, \\
\label{Oe} \textit{O}(e_{\alpha}, d^C_{\beta})&=& c(e_{\alpha},
d^C_{\beta}) \ \epsilon_{ijk} \ \overline{u^C_i} \ \gamma^{\mu} \
u_j \ \overline{d^C_{k \beta}} \
\gamma_{\mu} \ e_{\alpha},\\
\label{On} \textit{O}(\nu_l, d_{\alpha}, d^C_{\beta} )&=& c(\nu_l,
d_{\alpha}, d^C_{\beta}) \ \epsilon_{ijk} \ \overline{u^C_i} \
\gamma^{\mu} \ d_{j \alpha}
\ \overline{d^C_{k \beta}} \ \gamma_{\mu} \ \nu_l, \\
\label{OnC} \textit{O}(\nu_l^C, d_{\alpha}, d^C_{\beta} )&=&
c(\nu_l^C, d_{\alpha}, d^C_{\beta}) \ \epsilon_{ijk} \
\overline{d_{i \beta}^C} \ \gamma^{\mu} \ u_j \ \overline{\nu_l^C}
\ \gamma_{\mu} \ d_{k \alpha},
\end{eqnarray}
\end{subequations}
where the relevant coefficients are given by:
\begin{subequations}
\label{c}
\begin{eqnarray}
\label{cec} c(e^C_{\alpha}, d_{\beta})&=& k_1^2 [V^{11}_1 V^{\alpha
\beta}_2 + ( V_1 V_{UD})^{1
\beta}( V_2 V^{\dagger}_{UD})^{\alpha 1} ],\\
\label{ce} c(e_{\alpha}, d_{\beta}^C) &=& k^2_1  \ V^{11}_1
V^{\beta \alpha}_3 +  \ k_2^2 \
(V_4 V^{\dagger}_{UD} )^{\beta 1} ( V_1 V_{UD} V_4^{\dagger} V_3)^{1 \alpha},\\
\label{cnu} c(\nu_l, d_{\alpha}, d^C_{\beta})&=& k_1^2  ( V_1
V_{UD} )^{1 \alpha} ( V_3 V_{EN})^{\beta l} + k_2^2 V_4^{\beta
\alpha}( V_1 V_{UD}
V^{\dagger}_4 V_3 V_{EN})^{1l},\ \alpha=1 \ \textrm{or} \ \beta=1,\\
\label{cnuc} c(\nu_l^C, d_{\alpha}, d^C_{\beta})&=& k_2^2 [ ( V_4
V^{\dagger}_{UD} )^{\beta
 1} ( U^{\dagger}_{EN} V_2)^{l \alpha }+ V^{\beta \alpha}_4
 (U^{\dagger}_{EN} V_2 V^{\dagger}_{UD})^{l1}],  \ \ \alpha=1 \
 \textrm{or} \ \beta=1.
\end{eqnarray}
\end{subequations}
The mixing matrices $V_1= U_C^{\dagger} U$, $V_2=E_C^{\dagger}D$,
$V_3=D_C^{\dagger}E$, $V_4=D_C^{\dagger} D$,
$V_{UD}=U^{\dagger}D$, $V_{EN}=E^{\dagger}N$, and $U_{EN}=
E_C^{\dagger} N_C$. $\alpha,\beta=1,2$, $l=1,2,3$, while $i,j$,
and $k$ are the color indices. (Our convention for the
diagonalization of the up, down and charged lepton Yukawa matrices
is specified by $U^T_C Y_U U = Y_U^{\textrm{diag}}$, $ D^T_C Y_D D
= Y_D^{\textrm{diag}}$, and $E^T_C Y_E E = Y_E^{\textrm{diag}}$.)
The quark mixing is given by $V_{UD}=U^{\dagger}D=K_1 V_{CKM}
K_2$, where $K_1$ and $K_2$ are diagonal matrices containing three
and two phases, respectively. The leptonic mixing $V_{EN}=K_3
V^D_l K_4$ in case of Dirac neutrino, or $V_{EN}=K_3 V^M_l$ in the
Majorana case. $V^D_l$ and $V^M_l$ are the leptonic mixing
matrices at low scale in the Dirac and Majorana case,
respectively. The gauge $d=6$ operators have to be run from the
GUT scale down to 1\,GeV, i.e., the proton decay scale, and the
appropriate amplitude computed in the usual way. (For details, see
for example~\cite{Langacker}.)

In the above expressions $k_1= g_{GUT} M^{-1}_{(X,Y)}$, and $k_2=
g_{GUT} {M^{-1}_{(X',Y')}}$, where $M_{(X,Y)}$, $M_{(X',Y')}$
$\approx M_{GUT}$ are the masses of the superheavy gauge bosons.
All terms proportional to $k_1$ are obtained when we integrate out
$(X, Y)=({\bf 3},{\bf 2},5/3)$, where $X$ and $Y$ fields have
electric charge $4/3$ and $1/3$, respectively. These are the
fields appearing in theories based on the $SU(5)$ gauge group.
Thus, we call their contributions the ``$SU(5)$ contributions''.
Integrating out $(X', Y')=({\bf 3},{\bf 2},-1/3)$ we obtain the
terms proportional to $k_2$. These contributions we refer to as
the ``flipped $SU(5)$ contributions'' since they appear in the
flipped $SU(5)$ scenario. The electric charge of $Y'$ is $-2/3$,
while $X'$ has the same charge as $Y$. Again, there are no other
gauge contributions in any GUT besides these.

Minimization of the total decay rate represents formidable task
since there are in principle 42 unknown parameters. To face the
challenge we look for a solution where the ``$SU(5)$
contributions'' and ``flipped $SU(5)$ contributions'' are
suppressed (minimized) independently. Since we expect that in
general the associated gauge bosons and couplings have different
values this is also the most natural way to look for the minimal
decay rate value. Moreover, the bounds obtained is such a manner
will be independent of the underlying gauge symmetry.

The ``flipped $SU(5)$ contributions'' are set to zero by the
following two conditions~\cite{Ilja2}:
\begin{eqnarray}
V_4^{\beta \alpha}= (D_C^{\dagger} D)^{\beta \alpha}&=&0, \
\alpha=1 \ \textrm{or} \ \beta=1, \ \ (\textrm{Condition I}) \nonumber\\
(U_C^\dagger E)^{1\alpha}&=&0. \ \ (\textrm{Condition II}) \nonumber
\end{eqnarray}
(Condition I cannot be satisfied in the case of symmetric down
quark Yukawa couplings.) Therefore, in the presence of all gauge
$d=6$ contributions, in the Majorana neutrino case, there only
remain the contributions appearing in $SU(5)$ models. These,
however, cannot be set to zero~\cite{Nandi} in the case of three
generations of matter fields. But, as we now show, they can be
significantly suppressed. There are two major scenarios to be
considered that defer by the way proton decays:

\begin{itemize}

\item \textbf{There are no decays into the meson-charged
antilepton pairs}

All contributions to the decay of the proton into charged
antileptons and a meson can be set to zero . Namely, after we
implement Conditions I and II, we can set to zero Eq.~\eqref{ce}
by choosing
\begin{equation}
V_1^{11}= (U_C^\dagger U)^{11}=0. \ \ (\textrm{Condition III})
\end{equation}
(This condition cannot be implemented in the case of symmetric
up-quark Yukawa couplings.) On the other hand, Eq.~\eqref{cec} can
be set to zero only if we impose
\begin{equation}
(V_2 V_{UD}^\dagger)^{\alpha1 }= (E_C^\dagger U)^{\alpha 1}=0. \ \
(\textrm{Condition IV})
\end{equation}
Therefore with Conditions I--IV there are only decays into
antineutrinos and, in the Majorana neutrino case, the only
non-zero coefficients are:
\begin{equation}
\label{cnu1new} c(\nu_l, d_{\alpha}, d^C_{\beta})= k_1^2 \ ( V_1
V_{UD} )^{1 \alpha} ( V_3 V_{EN})^{\beta l}.
\end{equation}
So, indeed, there exists a large class of models for fermion
masses where there are no decays into a meson and charged
antileptons.

\noindent Up to this point all conditions we impose are consistent
with the unitarity constraint and experimental data on fermion
mixing. (In the $SU(5)$ case we have to impose Conditions III and
IV only.) We now proceed and investigate the decay channels with
antineutrinos. From Eq.~\eqref{cnu1new} we see that it is not
possible to set to zero all decays since the factor $( V_1 V_{UD}
)^{1 \alpha}$ can be set to zero for only one value of $\alpha$ in
order to satisfy the unitarity constraint. Therefore we have to
compare the following two cases:

\begin{itemize}

\item \textbf{Case a)} $( V_1 V_{UD} )^{1 1}=0$ (Condition V).

In this case the chiral langragian technique yields:
\begin{eqnarray*}
\Gamma_a(p \to \pi^{+} \overline{\nu}_i)&=&0,\\
\Gamma_a(p \to K^{+} \bar{\nu})&=& C(p,K) \left[1 + \frac{m_p}{3
m_B} (D + 3F) \right]^2 \frac{|V_{CKM}^{32} V_{CKM}^{21} -
V_{CKM}^{31} V_{CKM}^{22}|^2}
{|V_{CKM}^{31}|^2 + |V_{CKM}^{21}|^2},\\
\Gamma_a(n \to \pi^{0} \overline{\nu}_i)&=&0,\\
\Gamma_a(n \to K^{0} \bar{\nu})&=& C(n,K) \left[ 1 + \frac{m_n}{3
m_B} (D + 3F) \right]^2 \frac{|V_{CKM}^{32} V_{CKM}^{21} -
V_{CKM}^{31} V_{CKM}^{22}|^2}
{|V_{CKM}^{31}|^2 + |V_{CKM}^{21}|^2},\\
\Gamma_a(n \to \eta \overline{\nu}_i)&=&0,
\end{eqnarray*}
where:
\begin{equation}
C(a,b)= \frac{(m_a^2 - m_b^2)^2}{8 \pi m_a^3 f^2_{\pi}} \ A_L^2 \
|\alpha|^2 \ k_1^4.
\end{equation}

\item \textbf{Case b)} $( V_1 V_{UD} )^{1 2}=0$ (Condition VI).

All the decays channels into antineutrinos are non-zero in this
case. Associated decay rates are:
\begin{eqnarray*}
\Gamma_b(p \to \pi^{+} \bar{\nu})&=& C(p,\pi) \left[1 + D +
F\right]^2 \frac{|V_{CKM}^{32} V_{CKM}^{21} - V_{CKM}^{31}
V_{CKM}^{22}|^2}
{|V_{CKM}^{22}|^2 + |V_{CKM}^{32}|^2},\\
\Gamma_b(p \to K^{+} \bar{\nu})&=& C(p,K) \left[\frac{2 m_p}{3
m_B} D\right]^2 \frac{|V_{CKM}^{32} V_{CKM}^{21} - V_{CKM}^{31}
V_{CKM}^{22}|^2}
{|V_{CKM}^{22}|^2 + |V_{CKM}^{32}|^2},\\
\Gamma_b(n \to \pi^0 \bar{\nu})&=&
C(n,\pi) \frac{[1 + D +
F]^2}{2} \Gamma(p \to \pi^{+} \bar{\nu}),\\
\Gamma_b(n \to K^0 \bar{\nu})&=& C(n,K) \left[ 1 + \frac{m_n}{3
m_B} (D - 3F) \right]^2 \frac{|V_{CKM}^{32} V_{CKM}^{21} -
V_{CKM}^{31} V_{CKM}^{22}|^2}
{|V_{CKM}^{22}|^2 + |V_{CKM}^{32}|^2},\\
\Gamma_b(n \to \eta \overline\nu)&=& C(n,\eta) \frac{[1+D-3
F]^2}{6} \frac{|V_{CKM}^{32} V_{CKM}^{21} - V_{CKM}^{31}
V_{CKM}^{22}|^2} {|V_{CKM}^{22}|^2 + |V_{CKM}^{32}|^2}.\nonumber \\
\end{eqnarray*}

\end{itemize}

The nice thing about these results is that they are completely
independent of \textit{all}\/ CP violating phases including those
of $V_{CKM}$ and $V_l$ and any mixing angles beyond the $CKM$
ones. (This is completely unexpected since there are in principle
42 different angles and phases that could \textit{a priori}\/
enter our results.) Also, in the limit $V_{CMK}^{13} \rightarrow
0$ all decay rates vanish as required in the case of three
generations of matter fields~\cite{Nandi}. To demonstrate these
two properties we adopt the so-called ``standard'' parametrization
of
$V_{CKM}$~\cite{Chau:1984fp,Harari:1986xf,Fritzsch:1986gv,Botella:1985gb}
that utilizes angles $\theta_{12}$, $\theta_{23}$, $\theta_{13}$,
and a phase $\delta_{13}$. (For example, in that parametrization
$V_{CKM}^{13}=e^{-i \delta_{13}} s_{13}$.) The relevant terms read
$V_{CKM}^{32} V_{CKM}^{21} - V_{CKM}^{31} V_{CKM}^{22}=e^{i
\delta_{13}} s_{13}$, $|V_{CKM}^{22}|^2 +
|V_{CKM}^{32}|^2=c_{12}^2 + s_{12}^2 s_{13}^2$ and
$|V_{CKM}^{31}|^2 + |V_{CKM}^{21}|^2=s_{12}^2 + c_{12}^2
s_{13}^2$, where $c_{ij}=\cos \theta_{ij}$ and $s_{ij}=\sin
\theta_{ij}$. Hence, all one needs to know are angles
$\theta_{12}$ and $\theta_{13}$.

We present numerical values of all relevant two body decay
lifetimes for proton and bounded neutron decays in
Tables~\ref{tab:table1} and \ref{tab:table2}, respectively.
Clearly, it is \textbf{Case b)} that gives the lowest total decay
rate in the Majorana neutrino case. (We also include the Dirac
neutrino case for completeness.) Lifetimes are given in units of
$M^4_X/\alpha_{GUT}^2$, where the gauge boson mass is taken to be
$10^{16}$\,GeV. To generate these values we use $m_p=938.3$\,MeV,
$D=0.81$, $F=0.44$, $m_B=1150$\,MeV, $f_{\pi}=139$\,MeV,
$A_L=1.43$, and the most conservative value
$\alpha=0.003$\,GeV$^3$~\cite{Aoki}. Indicated uncertainties
reflect the errors in measurement of angles $\theta_{12}$ and
$\theta_{13}$ only. These are well-known and their sines are:
$s_{12}=0.2243 \pm 0.0016$, and $s_{13}=0.0037 \pm
0.0005$~\cite{PDG2004}. Note that the most poorly known parameter
is actually $\alpha$; the most recent QCD lattice
calculations~\cite{heplat0402026,heplat0409114} indicate that its
value could be three times bigger than the value we use. If that
result persists it would reduce the lifetime bounds we present by
a factor of ten.

\begin{table}[h]
\caption{\label{tab:table1}Proton lifetimes in years for Majorana
and Dirac neutrinos in units of $M^4_X/\alpha_{GUT}^2$, where the
mass of gauge bosons is taken to be $10^{16}$\,GeV.}
\begin{ruledtabular}
\begin{tabular}{lcccc}
 &\multicolumn{2}{c}{Majorana}&\multicolumn{2}{c}{Dirac}\\
 Channel&\textbf{Case a)}&\textbf{Case b)}&\textbf{Case a)}
&\textbf{Case b)}\\ \hline $p\rightarrow{{\pi }^+}\bar{\nu }$ &
$\infty$ & $5.1^{+1.7}_{-1.1} \times {{10}^{38}}$ &
$5.4^{+1.8}_{-1.2} \times {{10}^{38}}$ &
$2.6^{+0.9}_{-0.6} \times{{10}^{38}}$ \\
$p\rightarrow{K^+}\bar{\nu}$ & $1.0^{+0.4}_{-0.2} \times
{{10}^{38}}$ & $2.5^{+0.9}_{-0.6} \times {{10}^{40}}$ &
$6.8^{+0.0}_{-0.0} \times {{10}^{36}}$ & $7.2^{+0.0}_{-0.0} \times{{10}^{36}}$\\
TOTAL & $1.0^{+0.4}_{-0.2} \times{{10}^{38}}$ & $5.0^{+1.7}_{-1.1}
\times{{10}^{38}}$ & $6.7^{+0.0}_{-0.0} \times{{10}^{36}}$ &
$7.1^{+0.0}_{-0.0} \times{{10}^{36}}$\\
\end{tabular}
\end{ruledtabular}
\end{table}

\begin{table}[h]
\caption{\label{tab:table2}Lifetimes for bounded neutrons in years
for Majorana and Dirac neutrinos in units of
$M^4_X/\alpha_{GUT}^2$, where the mass of gauge bosons is taken to
be $10^{16}$\,GeV.}
\begin{ruledtabular}
\begin{tabular}{lcccc}
 &\multicolumn{2}{c}{Majorana}&\multicolumn{2}{c}{Dirac}\\
 Channel&\textbf{Case a)}&\textbf{Case b)}&\textbf{Case a)}
&\textbf{Case b)}\\ \hline $n \rightarrow \pi^0\bar{\nu}$
&$\infty$ & $1.0^{+0.3}_{-0.2} \times{{10}^{39}}$ &
$1.1^{+0.4}_{-0.2} \times{{10}^{39}}$&$5.2^{+1.8}_{-1.2} \times{{10}^{38}}$\\
$n\rightarrow{K^0}\bar{\nu}$ & $1.1^{+0.4}_{-0.2}
\times{{10}^{38}}$ &$ 6.7^{+2.3}_{-1.5} \times{{10}^{39}}$ &
$1.9^{+0.0}_{-0.0} \times{{10}^{36}}$ &
$1.9^{+0.0}_{-0.0} \times{{10}^{36}}$\\
$n\rightarrow\eta \bar{\nu}$ & $\infty$ & $1.5^{+0.5}_{-0.3}
\times{{10}^{41}}$ &$1.6^{+0.5}_{-0.3} \times{{10}^{41}}$ &
$7.6^{+2.5}_{-1.7} \times{{10}^{40}}$\\
TOTAL & $1.1^{+0.4}_{-0.2} \times{{10}^{38}}$ &$8.8^{+2.9}_{-2.0}
\times{{10}^{38}}$ &
$1.9^{+0.0}_{-0.0} \times{{10}^{36}}$ &$1.9^{+0.0}_{-0.0} \times{{10}^{36}}$\\
\end{tabular}
\end{ruledtabular}
\end{table}
\newpage
\item \textbf{There are no decays into the meson-antineutrino pair
in the Majorana neutrino case}

Let us show that it is also possible to set to zero all nucleon
decay channels into a meson and antineutrinos. After Conditions I
and II, we could impose $(V_1 V_{UD})^{1 \alpha}=0$ (Condition
VII) instead of $V_1^{11}=0$. (Again, these two equalities are
exclusive in the case $V^{13}_{CKM} \neq 0$.) Therefore, in the
Majorana neutrino case, there are no decays into antineutrinos
(see Eq.~\eqref{cnu}). In this case the property that the gauge
contributions vanish as $|V^{13}_{CKM}| \rightarrow 0$ is obvious
since $|V_1^{11}|=|V^{13}_{CKM}|$. We have to further investigate
all possible values of $V_2^{\beta \alpha}$ and $V_3^{\beta
\alpha}$. Now, we can choose $V_2^{\beta \alpha}=0$ and
$V_3^{\beta \alpha}=0$, except for the case $\alpha=\beta=2$
(Condition VIII). In that case there are only decays into a
strange mesons and muons. Let us call this \textbf{Case c)}. To
understand which case gives us an upper bound on the total proton
decay lifetime in the Majorana neutrino case, we compare the
predictions coming from the \textbf{Case b)} and \textbf{Case c)}.
The ratio between the relevant decay rates is given by:
\begin{equation}
\frac{\Gamma_c(p \to K^0 \mu^+)}{\Gamma_b(p \to \pi^+ \bar{\nu})}=
2 (c_{12}^2 + s_{12}^2 s_{13}^2) \frac{(m_p^2 - m_K^2)^2}{(m_p^2 -
m_{\pi}^2)^2}\frac{[1 + \frac{m_p}{m_B}(D-F)]^2}{[1+D+F]^2}=0.33
\end{equation}
Thus, the upper bound on the proton lifetime in the case of
Majorana neutrinos indeed corresponds to the total lifetime of
\textbf{Case c)}. We find it to be:
\begin{equation}
\tau_p\leq 1.5^{+0.5}_{-0.3} \times 10^{39} \
\frac{(M_X/10^{16}\,\textrm{GeV})^4}{\alpha_{GUT}^2} \
(0.003\,\textrm{GeV}^3 / \alpha)^2\,\textrm{years},
\end{equation}
where the gauge boson mass is given in units of $10^{16}$\,GeV. We
explicitly indicate the dependence of our results on the nucleon
decay matrix element. These bounds are applicable to any GUT
regardless whether the scenario is supersymmetric or not. If the
theory is based on $SU(5)$ the above bounds are obtained by
imposing Conditions VII and VIII. If the theory contains both
$SU(5)$ and flipped $SU(5)$ contributions, in addition to these,
one needs to impose Conditions I and II.

\end{itemize}

\begin{figure}[h]
\begin{center}
\includegraphics[width=4.5in]{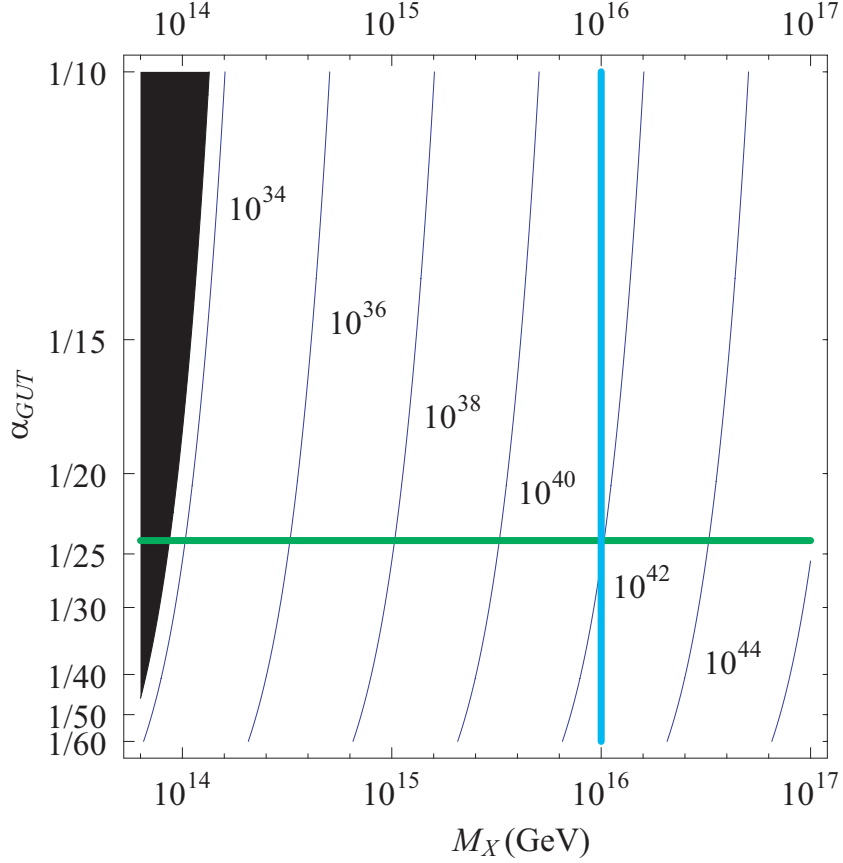}
\end{center}
\caption{\label{figure1} Isoplot for the upper bounds on the total
proton lifetime in years in the Majorana neutrino case in the
$M_X$--$\alpha_{GUT}$ plane. The value of the unifying coupling
constant is varied from $1/60$ to $1/10$. The conventional values
for $M_X$ and $\alpha_{GUT}$ in SUSY GUTs are marked in thick
lines. Experimentally excluded region is given in black.}
\end{figure}

\begin{figure}[h]
\begin{center}
\includegraphics[width=4.5in]{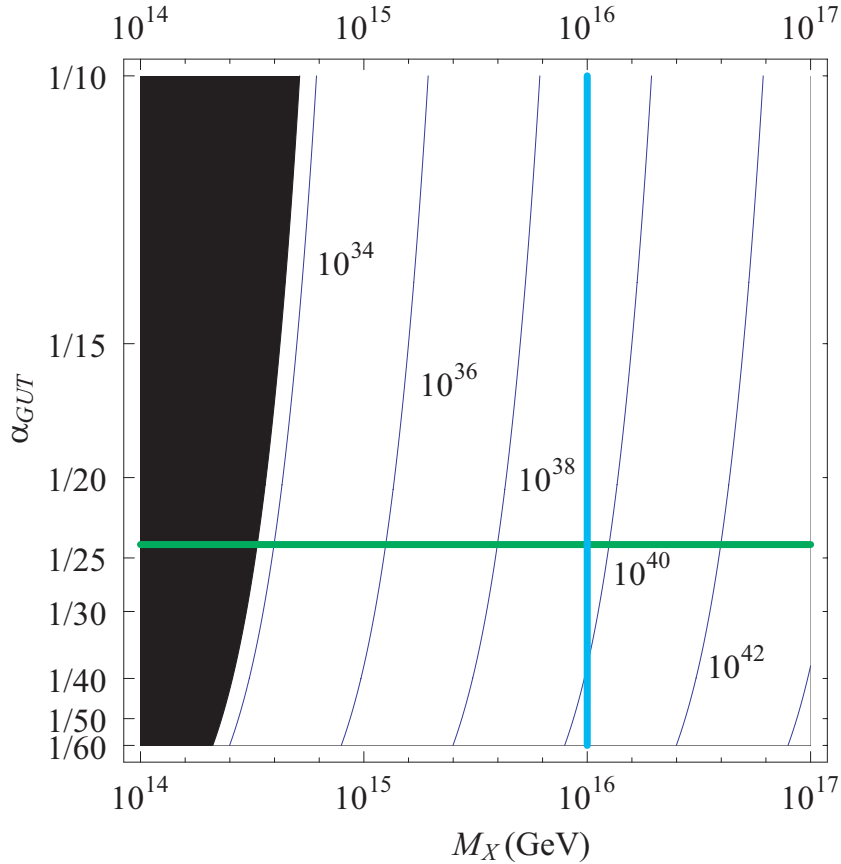}
\end{center}
\caption{\label{figure2} Isoplot for the upper bounds on the total
proton lifetime in years in the Dirac neutrino case in the
$M_X$--$\alpha_{GUT}$ plane. The value of the unifying coupling
constant is varied from $1/60$ to $1/10$. The conventional values
for $M_X$ and $\alpha_{GUT}$ in SUSY GUTs are marked in thick
lines. Experimentally excluded region is given in black.}
\end{figure}

We plot the proton bounds in the $M_X$--$\alpha_{GUT}$ plane for
the Majorana (Dirac) neutrino case in Fig.~\ref{figure1}
(\ref{figure2}). Our results, in conjunction with the experimental
limits on nucleon lifetime, set an absolute lower bound on mass of
superheavy gauge bosons. Since their mass is identified with the
unification scale after the threshold corrections are incorporated
in the running~\cite{Hall} this also sets the lower bound on the
unification scale. Using the most stringent limit on partial
proton lifetime ($\tau_p \geq 50 \times 10^{32}$\,years) for the
$p \rightarrow \pi^0 e^+$ channel~\cite{PDG2004} and setting
$\alpha=0.003$\,GeV$^3$~\cite{Aoki}, we obtain
\begin{equation}
M_X \geq 4.3^{+0.3}_{-0.3} \times 10^{14}
\sqrt{\alpha_{GUT}}\,\textrm{GeV},
\end{equation}
where $\alpha_{GUT}$ \textit{usually}\/ varies from $1/40$ for
non-supersymmetric theories to $1/24$ for supersymmetric theories.
For example, if we take a non-supersymmetric value
$\alpha_{GUT}=1/39$, we obtain
\begin{equation}
\label{GUT} M_X \geq 7 \times 10^{13}\,\textrm{GeV}.
\end{equation}
Again, this result implies that any
non-supersymmetric theory with $\alpha_{GUT}=1/39$ is eliminated
if its unifying scale is bellow $7.0 \times 10^{13}$\,GeV
regardless of the exact form of the Yukawa sector of the theory.
Note that majority of non-supersymmetric extensions of the
Georgi-Glashow $SU(5)$ model yield GUT scale which is slightly
above $10^{14}$\,GeV. Hence, as far as the experimental limits on
proton decay are concerned, these extensions still represent
viable scenarios of models beyond the SM. Region of $M_X$ excluded
by the experimental result is also shown in Figs.~\ref{figure1}
and~\ref{figure2}.

At this point the following two observations are in order:
\begin{enumerate}
    \item All three cases (\textbf{Case a)}--\textbf{c)}) yield
comparable lifetimes (within a factor of ten) even though they
significantly defer in decay pattern predictions;
    \item We use the most stringent experimental limit on partial
proton lifetime as if it represents the limit on the total proton
lifetime. Even though this is not correct (see discussion
in~\cite{PDG2004}) it certainly yields the most conservative bound
on $M_X$.
\end{enumerate}

One can easily extend our results to a class of orbifold GUT
theories~\cite{Kawamura:1999nj,Kawamura:2000ev} where all matter
fields live on an ``unbroken'' brane. In essence, to obtain the
lower limit on the gauge boson mass in those theories, it suffices
to multiply the limit presented in Eq.~\eqref{GUT} by
$\sqrt{\pi/2}$. (This factor accounts for the fact that the
two-body decay of the proton is due to exchange of an entire
Kaluza-Klein (KK) tower of states~\cite{Hebecker} associated with
the proton decay mediating gauge boson.) The bound obtained in
such a way then corresponds to the limit on the compactification
scale of extra dimension(s). (Recall that in orbifold GUTs the
gauge bosons responsible for proton decay belong to the KK tower
where the lightest gauge boson in the tower has the mass equal to
the orbifold compactification scale.) Curiously enough, exact
unification of gauge couplings in the five-dimensional $S^1/(Z_2
\times Z'_2)$-type orbifold models usually requires the
compactification scale to be slightly above
$10^{14}$\,GeV~\cite{Hall:2001xb,Kim:2002im,Dorsner:2003yg}. This
would imply that the orbifold GUT theories with the matter fields
all located on the ``unbroken'' brane could soon be completely
ruled out if the proton decay is not observed in the next
generation of the proton decay experiments.

In order to complete our analysis let us finally demonstrate the
possibility to set to zero the Higgs $d=6$ and $d=5$
contributions. The triplets $T=({\bm 3},{\bm 1},-2/3)$ and
$\bar{T}=(\bar{{\bm 3}},{\bm 1},2/3)$ have the following
interactions:
\begin{equation}
W_T= \int d^2 \theta \left\{ \ [ \ \hat{Q} \ \underline{A} \ \hat{Q} \
  + \ \hat{U^C} \
\underline{C} \ \hat{E^C} \ + \ \hat{D^C} \ \underline{E} \ \hat{N^C}
\ ] \ \hat{T} \ + \ [ \ \hat{Q} \ \underline{B} \ \hat{L} + \hat{U^C} \
\underline{D} \ \hat{D^C} \ ] \ \hat{\bar{T}} \ \right\} + \ \text{h.c.}
\end{equation}
Choosing $\underline{A}_{ij}=-\underline{A}_{ji}$ and
$\underline{D}_{ij}=0$, except for $i=j =3$, the Higgs $d=6$ and
$d=5$ contributions are indeed set to zero. It is also possible to
have SUSY scenarios where the d=5 operators are strongly
suppressed by particular realization of superparticle
spectrum~\cite{Arkani}. In any case, even if SUSY is realized at
low energies we are sure that the upper bound is coming from the
gauge $d=6$ contributions.

\section{Summary}

We have investigated the possibility of finding an upper bound on
the total nucleon decay lifetime in the context of grand unified
theories with the Standard Model matter content. This bound
originates from the gauge $d=6$ contributions, since all other
contributions are quite model dependent and can always be
suppressed. In the Majorana neutrino case the bound is $\tau_p\leq
1.5 \times 10^{39}
\frac{(M_X/10^{16}\,\textrm{GeV})^4}{\alpha_{GUT}^2}
(0.003\,\textrm{GeV}^3 / \alpha)^2\,\textrm{years}$, while in the
Dirac neutrino case $\tau_p \leq 7.1 \times 10^{36}
\frac{(M_X/10^{16}\,\textrm{GeV})^4}{\alpha_{GUT}^2}
(0.003\,\textrm{GeV}^3 / \alpha)^2\,\textrm{years}$. These bounds
are valid in both supersymmetric and non-supersymmetric scenarios
and are grand unifying gauge group independent. Moreover, there is
no dependence of our results on CP violating phases nor any angles
beyond those of $CKM$. Our bounds are very useful for two reasons.
Firstly, in the context of realistic grand unified theories they
indicate whether it is possible to test these theories in their
\textit{entire}\/ flavor parameter space with certainty through
proton decay experiments. Secondly, they put an absolute lower
bound on the mass of proton decay mediating gauge bosons. We
obtain $M_X \geq 4.3^{+0.3}_{-0.3} \times 10^{14}
\sqrt{\alpha_{GUT}}$\,GeV for a reasonable set of input
parameters. Since this mass is usually identified with the
unifying scale through threshold matching conditions our bounds
can be interpreted as the lower bounds on the GUT scale itself. We
have also addressed implications our bounds have on the popular
class of the so-called ``orbifold'' models.

\begin{acknowledgments}
We would like to thank Bobby Acharya, Borut Bajc, Marco Aurelio Diaz, 
and Goran Senjanovi\'c for discussions and comments. 
P.F.P thanks the High Energy Section of the ICTP 
for their hospitality and support. This work was supported in part
by CONICYT/FONDECYT under contract $N^{\underline 0} \ 3050068$.
\end{acknowledgments}


\end{document}